# User interface design for military AR applications

Mark A. Livingston · Zhuming Ai ·
Kevin Karsch · Gregory O. Gibson

**Abstract** Designing a user interface for military situation awareness presents challenges for managing information in a useful and usable manner. We present an integrated set of functions for the presentation of and interaction with information for a mobile augmented reality application for military applications. Our research has concentrated on four areas. We filter information based on relevance to the user (in turn based on location), evaluate methods for presenting information that represents entities occluded from the user's view, enable interaction through a top-down map view metaphor akin to current techniques used in the military, and facilitate collaboration with other mobile users and/or a command center. In addition, we refined the user interface architecture to conform to requirements from subject matter experts. We discuss the lessons learned in our work and directions for future research.

## 1 Introduction

Military operations in urban terrain (MOUT) require creative solutions to overcome fundamental difficulties faced by tactical leaders. Military personnel engaged in both combat and non-combat operations must understand a complex, dynamic environment, of which they often see only a small portion. This understanding should be customized so that each user sees exactly what he needs to know—no more and no less. The ability to change plans during an operation while maintaining situation awareness (SA) between small, dispersed units is an important and new requirement in recent operations.

One of the main considerations to making such a system be both usable and useful is the design of the user interface (UI). Key goals include the intuitive and focused display of information and a natural way to interact with that information. By intuitive, we mean metaphors for presentation that are easy to understand and integrate with the 3D environment and the user's current understanding of that environment. By focused, we mean that the amount of information displayed is sufficient for the user to maintain situation awareness, but not so great that information is lost or obscured by other information. By natural, we mean interactions that are compatible with existing military information presentation and control, as well as being compatible with the other tasks expected from military personnel. These interactions must occur between multiple mobile systems and between mobile systems and command applications in fixed facilities.

In this manuscript, we focus on the methods of presenting, organizing, and interacting with information presented in a mobile augmented reality (AR) prototype for dismounted military users. The integration of, automation of, and interaction with the building blocks of our UI was an important step in moving our research program to an application prototype. We discuss our application context, feedback from domain experts, the integrated system design, and an initial evaluation of one aspect of the information presentation interface.

### 1.1 Related work

The Touring Machine (Feiner et al. 1997) introduced several visual representations fundamental to SA. The UI

M. A. Livingston (✉) · Z. Ai · K. Karsch · G. O. Gibson
Naval Research Laboratory, Washington, DC, USA
e-mail: mark.livingston@nrl.navy.mil

was built on menus that could be engaged through a handheld computer or through a see-through display. The software placed virtual labels on real buildings, which were selected by keeping them in the center of the display for one full second. Selection invoked additional menus which could be used to retrieve further information about that building. A compass pointer assisted the user in keeping a building in view and was especially useful for buildings selected via the hand-held computer's menu.

The MARS project (Höllerer et al. 1999) extended the UI to four possible configurations. The outdoor options included a head-worn display and a hand-held display; indoor options included desktop and immersive variations. For the outdoor UI, they separated objects into screen-fixed and world-fixed objects. The former included traditional UI widgets such as menus and selection cursors. The latter could include any model or label registered to the 3D environment. (Other objects had hybrid fixations, helping relate the UI widgets to world objects being affected.)

Tinmith (Piekarski and Thomas 2002) incorporated a series of interaction tools to control and create virtual information within the surrounding environment. A glove-based series of gestures (tracked through a combination of a pinch detection and synthetic markers affixed to the glove) could be used to navigate menus, select graphical objects, or manipulate objects. Similarly, an eye cursor was also used for selection. Manipulation was restricted to the image plane, which reduced the complexity of the interface.

The design space for ubiquitous AR interfaces is quite large, and exploring the complete set of choices may not be feasible (Sandor and Klinker 2007). However, a disciplined approach for a specific application can help in analyzing the options available. In our case, we employed structured formative evaluations with subject matter experts.

### 1.2 Application context

Over the course of an extended research program, we have conducted a series of interviews with subject matter experts (SMEs), both from the proposed military user community (officers with experience in dismounted infantry combat techniques) and from the military information management field. Three interview sessions were conducted. The first was a 1-h demonstration and discussion with a reserve officer with recent combat experience; this demonstration was of our previous system (Livingston et al. 2006), and the interview focused on the SME's opinions on how such a system might be useful and what information might be helpful for certain tasks. The second interview consisted of a day-long discussion with a recently retired combat officer, while the second was a half-day session with a panel of active duty dismounted infantry officers with combat experience; between eight and ten were in the room at any given time. Some discussions concentrated on general principles and designs for the application, and some focused on particular aspects of the interface. To assist the military users in understanding the capabilities of AR, in the second and third sessions, we showed concept sketches and snapshots from a prototype of the system; however, in these sessions, we did not have a completed system in which they could look through the HMD. Discussions with the military information management experts consisted of a series of discussions and emails over the course of the project and focused on assumptions that could be made about available information and what information potentially coordinated systems would make available. In this section, we discuss general guidelines elicited and/or confirmed in our most recent interviews; discussion of particular features in the user interface appears in the sections dealing with those features.

The first general question we asked of our military SMEs was what tasks they felt would benefit the most from an AR system. While we received a variety of answers, the most important (and consistent with previous interviews with other SMEs) piece of information was that knowing the locations of other friendly forces operating in the area would be extremely helpful. This is among the most fundamental aspects of SA in military operations (Bolstad and Endsley 2002). Other basic information, including building and street labels and a compass for knowing current orientation (relative to either a global or local coordinate frame), was also valued by the SMEs. Other suggestions were to incorporate route data (for the user and for other team members), event history data in recent days in the area of operations, rendezvous points, and objects tagged by other users in the system.

Our information management SMEs cautioned us against assuming that a precise model of the operating environment would be available prior to system initialization. Though this surprised us, we incorporated this constraint into our design process. A related constraint was imposed for outdoor AR tracking (Azuma et al. 2006). That research was primarily concerned with the registration requirements, but similarly eschewed a heavy infrastructure involving hundreds or thousands of tracking beacons.

A general reminder (also consistent with past interviews) was that the users' hands would often be occupied with existing military equipment or procedures. This argues for simple interface designs and maintaining consistency with existing tools. Another general reminder was the expectation that the system would be used in a high-stress environment, also arguing for UI simplicity, but also specifically leading us to develop the filtering algorithm described below. A recommendation we considered in the past, but was this time more strenuously suggested, was the use of military standard symbols (DISA 2008) to represent objects whenever possible. These function much as textual labels might, but convey information in ways that are familiar to the intended users of our application.

## 2 Information presentation for collaboration

Our domain analysis indicated several important requirements for leaders of small units (4–40 subordinates, depending on level in the hierarchy). One need was to focus only on information relevant to his team and its area of operation (which expands with the level of the hierarchy). In response, we developed an information filtering algorithm (Sect. 2.1) Another requirement was to be aware of the locations of friendly troops among urban infrastructure; this led us to investigate metaphors for depicting occluded objects or people (Sect. 2.2). Military users are accustomed to various implementations of maps (paper and electronic), and it was judged to be important to have this feature as part of our system (Sect. 2.3). Of course, enabling communication up-and-down the chain of command is always a critical element in military operations, so we provide a command-and-control console application (Sect. 2.4).

2.1 Information filtering

With AR, by tracking the user's position and orientation, complicated spatial information can be directly registered to the real world in the context where it applies. An urban combat environment is extremely complicated: the city is populated by large numbers of buildings, each of which can have numerous facts stored about it; friendly and hostile entities (people, vehicles) are constantly changing their positions. Therefore, it is very easy to cause the user to experience information overload. The display may include both relevant information and irrelevant information to a user's task.

To overcome these problems, we have developed algorithms for information filtering. These tools automatically restrict the information which is displayed to minimize problems of information overload. The approaches are based on the Concept of Operations derived from our SMEs interviews; they include modifications from a region-based information filter proposed previously (Julier et al. 2002). This new algorithm is a hybrid of the spatial model of interaction (Benford and Fahlen 1993), the rule-based filtering, and the definition of an operation zone.

The spatial model of interaction is a more sophisticated version of distance-based filtering. The spatial model was first developed to consider the problems of spatial awareness and interaction in multi-user virtual environments,
where awareness can be used to determine whether or not an object is visible to, or capable of interaction with, another object. In this model, each object (e.g., a user) is surrounded by a focus, specific to a medium, which defines the part of the environment of which the object is aware in that medium. Each object in the environment also has a medium-specific nimbus, which demarcates the space within which other objects can be aware of that object. The level of awareness that object A has of object B in medium M is some function of A's focus on B in M and B's nimbus on A in M. The spatial model has the advantage that it allows different objects to be demarcated at different ranges. The algorithm consists of the following steps.

1. Define an operation zone.
   In the mission preparation stage, a operation zone should be defined. It could be a patrol route (perhaps crossed by phase lines), defense area (demarcated by lines of deconfliction), target area of attack, etc. The operation may be modified during the mission.
2. User's focus.
   Each user has at least two foci and could in theory have a third. One is the range his firearm can cover, the other is an interactively defined range in which the user wants to be aware of information. He may wish to define a focus in the time dimension as well.
3. Calculate the impact zone for each object.
   An impact zone (nimbus) of an object is an extended region over which an object has a direct physical impact. An IED, for example, is effective over a larger distance if it is placed near a gas station. The impact zone can be represented as a sphere whose radius equals the maximum range of damage. Conversely, a more accurate representation could take account of the effects of buildings and terrain through modeling the impact zone as a series of interconnected volumes.
   The calculation of the impact zone is based on the properties of the object which include the object's classification (for example whether it is a mosque or a gas station), its location, its size, and its shape. The impact zone is also determined by the task and the intelligence and can be updated when new information comes. Examples include possible sniper coverage areas on a high building, possible explosion damage areas surrounding a gas station, etc. This calculation is carried out whenever an object's property changes or the user's objective changes.
4. Cull.
   Use the spatial model of interaction to determine which objects to hide and which to display. Those objects whose impact zone intersect with the operation zone are of interest. Those objects are not necessarily inside the operation zone. The following objects should be displayed, if

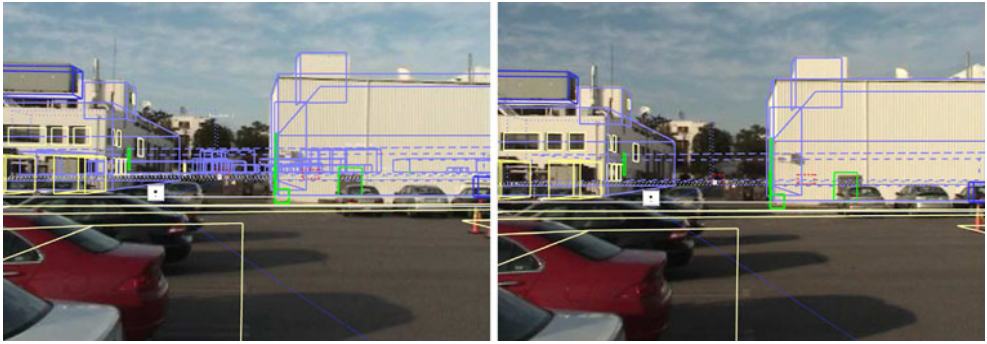

**Fig. 1** Examining the unfiltered view (*left*) and filtered view (*right*), it is easier to see the way-point icons (*center-left* and *center*) in the filtered view. Viewing through the HMD affords a larger angular size

- its nimbus intersects with the operation zone, and
  - it is within either of the user's foci, or
  - its nimbus intersects with the user's awareness focus.

This step is performed periodically when the user's position and/or orientation has changed.

The current implementation of the algorithm does not take the geometry of the buildings and terrain into account in calculating the impact zone, because such information could not be relied on in our applications. It is reasonable to assume that in the database we have the models (perhaps only 2D, perhaps with low fidelity or accuracy) of the objects that need to be displayed, such as buildings that might be used by snipers. Or these models can be represented by 3D icons that are designated on the scene by a user. However, according to our SMEs, we should not assume that we have a complete model of the environment.

In addition to this operation zone based filter, a rule-based filter ensures that all vital data, such as known enemy positions, IED positions, are always displayed. A filter manager initially sets the states of all the objects as "to be determined." The rule-based filter changes the state to "show" for vital information, and the operation zone based filter changes the state to "hide" for objects that are filtered out. The same filter manager makes occlusion representation part of the filtering system, a key point in integrating our information presentation system. This information filtering system worked well in our tests. The left image in Fig. 1 shows the view when no filter is applied; the right image is a much cleaner view when the filters are applied. The user's foci in the operation zone based filter can be adjusted interactively so that the amount of information displayed can be changed in real time. The user can adjust how much information he/she wants to see, the algorithm makes sure that the most important information is not missing from the view.

### 2.2 Occlusion

"X-ray vision"—the ability to see virtual representations of objects whose positions are occluded by the real environment, registered to that real environment—has long been cited as a desired feature in augmented reality (AR) applications (Furness 1969). The problem that has faced AR designers regarding this capability may be seen by examining Fig. 1. Superposition of the graphics, even assuming perfect registration, does not convey the depth of the graphical entities relative to the real objects visible in the environment. Though a number of visual metaphors have been designed to display such information, there are few comparisons of how well the various techniques work for users in an application context. Using the information learned from our SMEs [as well as our own previous study in this area (Livingston et al. 2003)], we designed and implemented a user study to compare several existing techniques, as well as one new variation that adapted existing techniques to the constraints outlined by the SMEs (specifically, not to assume the existence of a complete model and to use standard military symbols).

Our implementations focused on the use of simple graphics, mostly line drawings, to convey depth ordering or metric depth information. Using custom shader programs, we implemented six occlusion representation techniques (depicted in Fig. 2) in addition to a control condition with no changes in the representation based on the occlusion of the virtual objects by the real world.

1. *Opacity* We used a discretized function of distance to set the virtual object's opacity. This function maps more distant objects to lower opacity, which—when combined with a black rendering background—dims more distant objects, mimicking real-world behavior of distant objects.
2. *Stipple* Drawing inspiration from technical illustration, AR systems have used solid, dashed, and dotted lines to represent ordinal distance (Feiner and Seligmann

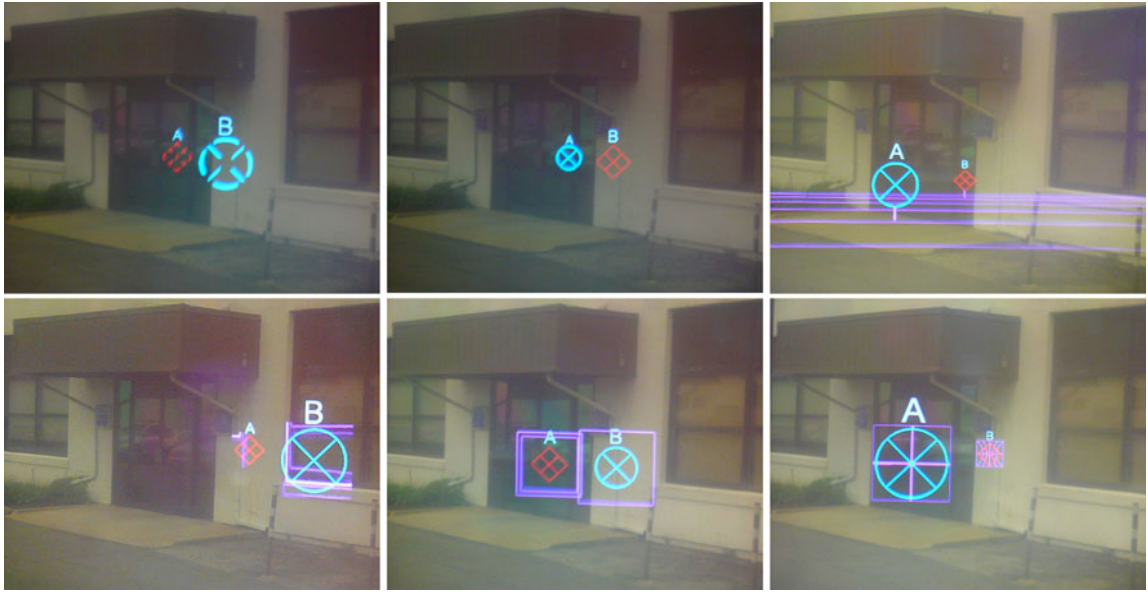

**Fig. 2** Examples of the occlusion metaphors as implemented in our user study. *Top row* (*left-to-right*): Stipple, Opacity, and Ground Grid. *Bottom row*: Edge Map, Tunnel, and Virtual Wall. Note that for the Ground Grid, the first visible element of the circular grid is 20 m from the user; all images were cropped identically in order to see the details in all the designs

1992). Our shader implementation fixed the stipple in object space, rather than screen space.

3. *Ground grid* A virtual ground plane can help the user understand relationships to flat terrain (Tsuda et al. 2006), building on the visual cues of relative size, texture gradient, and height in the visual field (Cutting 2003). Explicit markers that tie an object to the ground assist in this visualization.
4. *Edge map* Virtual representations of occluding edges convey the depth order between a real surface and occluded virtual objects (Avery et al. 2009). This metaphor benefits greatly from video-mediated AR (to acquire edges from video) and precise registration of the virtual objects.
5. *Virtual wall* A simple version of an edge map uses synthetic edges with the density of edges increasing with increasing ordinal depth to the virtual object behind the virtual wall.
6. *Virtual tunnel* The *virtual hole* metaphor (Bajura et al. 1992) extends to multiple surfaces, creating a virtual tunnel (Bane and Höllerer 2004). This technique works best for a single real surface, but with a model of the real environment, may apply to any number of virtual holes. Since we do not assume a complete model, we modify this technique to use squares to represent known occluding layers.

Fourteen (14) total subjects (11 male, 3 female) drawn from the research and clerical staff of our lab completed a study comparing these seven metaphors (including a control "Empty" condition).[1] Our volunteers were between age 20 and 48, received no compensation, and were heavy computer users (including five video game players). One other subject withdrew due to fatigue. Two subjects managed tracker errors by either waiting for it to subside (in one case) or simply ignoring it (in the other case) for the few trials in which it occurred. All users passed a stereo screening test. For each user, we calibrated IPD and height, then HMD orientation. The last of these could be repeated before any trial in the experiment, but users rarely felt the need to do so. Users stood in a position that gave rise to five depth "zones" (and were given a map showing these). They then attempted to correctly identify the zone for each of a pair of icons (Fig. 2). Other tasks and responses were performed and recorded, but we report only the results of this task for space considerations. The independent variable of interest was the occlusion metaphor; its presentation was counterbalanced with a Latin square.

As seen in Fig. 3, the Tunnel metaphor led to the lowest error, followed by the Virtual Wall and the Ground Grid. The Edge Map and the Empty design were the least helpful of the representations. This is the key result we sought in our experiment; it helps us identify which representations are worthy of further study and are most likely to be refined into a best method.

---

[1] In keeping with practice for psychophysical experiments, we did not attempt to find domain experts to serve as subjects. Depth perception and relative depth judgments do not require military experience to perceive.

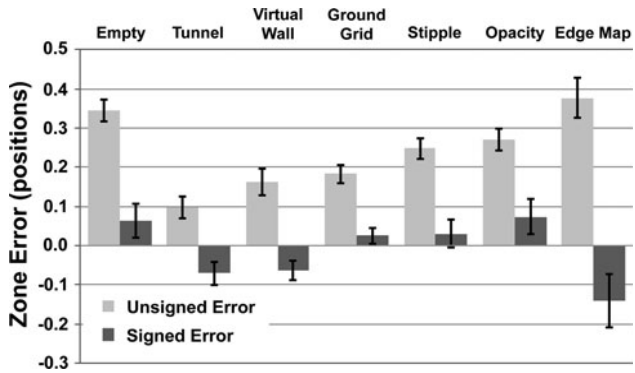

**Fig. 3** The graph of unsigned error (*light*) versus the occlusion metaphor shows that the Tunnel metaphor led to the least amount of error, followed by the Virtual Wall and the Ground Grid. The Edge Map led to the greatest amount of error, followed closely by no occlusion representation ("Empty"). Looking at the signed error (*dark*) shows that the Tunnel, Virtual Wall, and Edge Map led users to perceive the occluded object as closer than it really was (negative error)

We gain further insight into the performance with these candidate designs when we turn to the graph of signed error (also in Fig. 3), where a negative error indicates that the subjects perceived the icon to be closer than it really was. The Ground Grid metaphor had a signed error closest to zero. The Tunnel has a negative error. So while users tended to make the least errors with the Tunnel design, the direction of the errors that were made was in the negative direction; users perceived the icon correctly or as being closer than it was. The same can be said of the Virtual Wall; its error was also toward the negative. We also noted that users were fastest with the Empty design and the Virtual Tunnel design, while they were slowest with the Edge Map metaphor.

For the most part, the results of the study coincide with our intuition. The Empty and Edge Map metaphors provide the least additional information, resulting in a poor depth estimate. The Stipple and Opacity metaphors vary smoothly with distance, as does the size of the object for all metaphors. Thus, this information is somewhat redundant, but it is still better than no additional information. The Tunnel, Virtual Wall, and Ground Grid immediately narrow the range of depths that a user must consider, allowing for a quick and (more) accurate choice.

We asked users to indicate if they employed a particular strategy to solve the task. Four users indicated that they were trying to use the relative size of the object as a direct distance cue. One of these subjects conceived of the icon as having the height of a person; another tried to use a real object as a size cue.

The perceptual results from this pilot test will help us restrict the cases that we will present in a larger study once the application becomes adopted by our intended user community. We plan to eliminate the Edge Map and Empty designs as candidates. We also will explore the extensive parameter space on the remaining metaphors to enable us to compare the best implementation of each of those designs against each other.

### 2.3 Interaction through the map

In addition to the head-up view, the AR system has a map view mode (Feiner et al. 1997), which is automatically activated when the user looks down while wearing the HMD (as sensed by the orientation tracker affixed to the HMD). The military community relies heavily on maps for planning and coordinating collaborative operations, so this was an important feature for our target users. The map mode is implemented by moving the view point to a very high position and looking downwards, centering at the current user position and orienting according to the user's current orientation. This is not an AR view per se; instead it displays all the objects in the database from a bird's-eye view, shown on top of the ground at the user's feet. Since this usually limits the amount of background light, the AR graphics are generally visible in this merged image.

We use this mode not just as a visualization of the world from a traditional perspective, but also to provide an additional way for the user to interact with the system. Drawing in 3D can be very hard, especially when potential vertices correspond to locations that are hidden from the users's view. Despite the promise shown by the representations of occluded locations in the previous section, we have not attempted to implement a general 3D drawing interface for a mobile user. We have implemented a GUI through which the user can create and edit routes by drawing on the map. The mobile user interacts with the system using a gyroscopic mouse or a handheld trackball. (We have used the center of the field of view as the cursor in past implementations.) When editing a route, the user can add a way-point simply by a mouse click. The 2D mouse position is transformed to the 3D world coordinate by assuming the way-point is on the ground ($z = 0$). The way-points are connected to form a route. This feature has been tested and used to create routes for our demonstrations. It is analogous to clicking on an electronic map or drawing points on paper maps, as are done currently.

These routes are then considered by the filtering algorithm (Fig. 4, left). The map may also be used to preview the result of adjusting parameters of the filtering algorithm (Fig. 4, right). This global view has gotten better feedback from our domain experts than adjusting the filtering in the head-up AR view.

### 2.4 Command and control console

In collaborative missions, numerous sensors may collect data and relay that data to a command and control (C2)

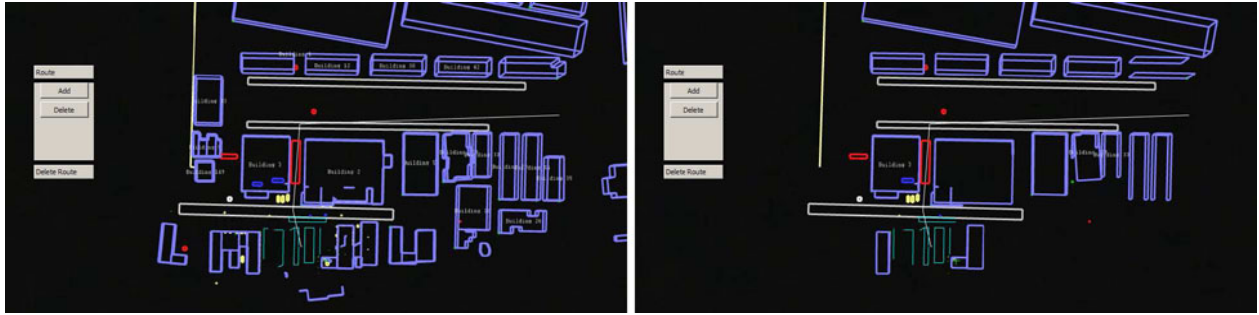

**Fig. 4** Map view is useful for editing routes (*left*) and previewing filter (*right*)

center. Integrating this geo-registered information becomes necessary to maintain SA of the environment. We have designed and implemented a system that integrates information from satellite/aerial images, 3D models, real-time video images, and other iconic information into a virtual globe application (Ai and Livingston 2009). Such an application enables a commander to view the environment from an arbitrary viewpoint, such as a bird's-eye view to get a globe understanding of 3D relationships between personnel and routes, or from a particular user's vantage, in order to work closely with that user on a specific task, such as altering a planned action to reduce risks.

We display multiple types of data in the C2 application. The basis for our implementation is the virtual global application, Google Earth. We also considered other virtual global applications such as NASA's World Wind. Google Earth is chosen because it has the features we need, such as a 3D building layer and API, to quickly develop a prototype. This enables us to use satellite imagery and simple 3D models extracted from such imagery as an approximate model of the world. If no such 3D models are available, then the satellite imagery (along with geographic terrain data) can serve as the basic model of the environment. We then add simple icons representing the positions of users or labeled objects in the environment. This is relatively simple once the tracking of such personnel or specification of locations (not in itself a simple problem) has been addressed. Finally, we project 2D imagery acquired from surveillance cameras onto the environment. These sensors may be fixed or tracked within the environment, such as by GPS or some other globally registered system. This is a more complex operation and requires that we have at least a rough model of the environment in order to have accurate projection matrices. We restrict the projection of images to the ground or to large structures such as buildings, rather than attempt to project onto vehicles or personnel.

To display the images on Google Earth correctly, we need to create the projected texture maps on the ground and the buildings. This requires the projected images and location and orientation of the texture maps. We create textures in the frame-buffer from the images with OpenSceneGraph and render them onto rectangles whose position and orientation are calculated from the camera's pose. When viewing from the camera position and using proper viewing and projection transformations, the needed texture maps are created by rendering the scene to the frame-buffer.

To create the texture map of the wall, an asymmetric perspective viewing volume is needed. The viewing direction is perpendicular to the wall. The viewing volume is a frustum which is formed with the camera position as the apex, and the wall (a rectangle) as the base. When projecting on the ground, we first divide the area of interest into grids of proper size. When each rectangular region of the grid is used instead of the wall, the same projection method for the wall described above is used to render the texture map in the frame-buffer. The zoom factor of the video camera is converted to the field of view. Together with the pose of the tracked camera, we calculate where to put the video images. The position and size of the image can be arbitrary, as long as it is along the camera viewing direction, with the right orientation and a proportional size. By integrating images, icons, and 3D models as shown in Fig. 5, it is very easy for the command and control center to monitor what is happening live on the ground.

## 3 Discussion

We redesigned and implemented the AR system to take advantage of the lessons we learned during years (Livingston et al. 2006) in developing AR systems intended for military use. We describe our new implementation and some lessons learned over the course of this project.

### 3.1 Implementation

The new system is written in C++; it currently runs on MS Windows systems. It is easy to port to other platforms, such as Linux, since all the development tools are multi-platform. The system supports the Lua scripting language[2] so that displayed objects and many parts of the user interface can easily be controlled by scripting; it is particularly useful in designing user studies (including the one described in this manuscript). An input/output module reads 3D models and other information into an internal database, which is shared among users and may be modified interactively by the user. The information then is sent through a serious of filters and generates a scene graph that is displayed by a rendering control module. The system is designed for cooperative missions, the High Level Architecture (HLA) is used to distribute information among users over the network. The Google Earth C2 component is also connected to the system via HLA which supports network video cameras as well as cameras connected directly to the computer. OpenSceneGraph[3] is used for graphics rendering, Delta3D[4] is used for synthetic force simulation, and Qt[5] is used for the GUI.

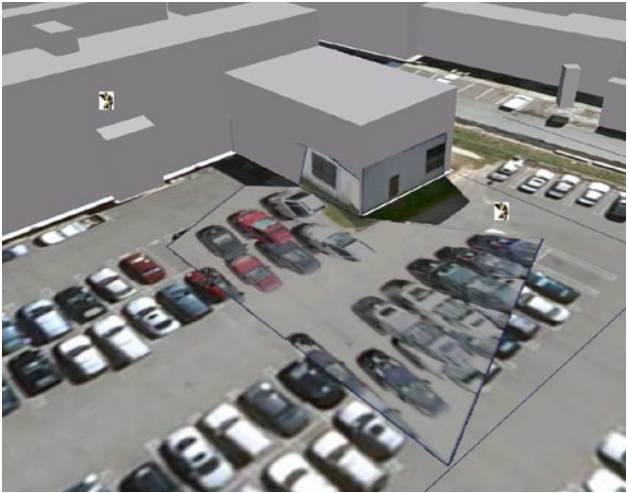

**Fig. 5** Recreated 3D scene viewed with 3D buildings on Google Earth. The two field operators' icons and the video image are overlaid on Google Earth

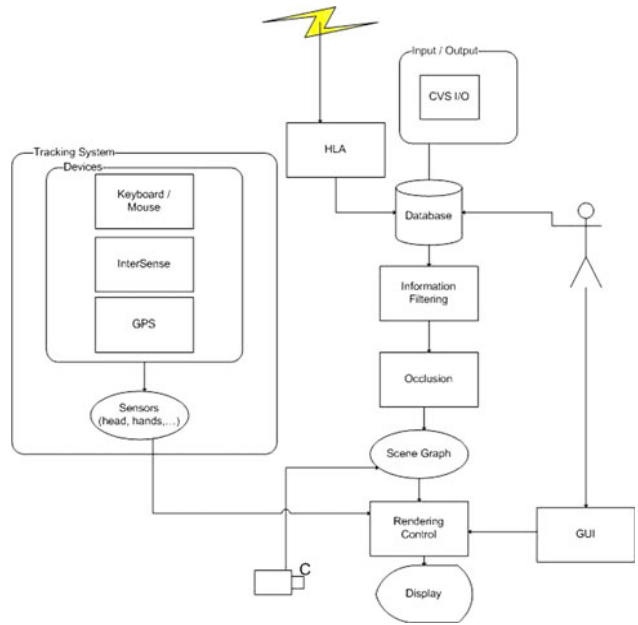

**Fig. 6** Software structure of the AR system

The system has a sensor control component that supports a variety of hardware and allows user to link a sensor (e.g., head position or orientation) to different hardware on the fly. Only commercial, off-the-shelf hardware products are used. The system can run on a mini-netbook with a head-mounted display (HMD).

### 3.2 Lessons learned

#### 3.2.1 UI architecture

Previously, we argued for a "mediator" architecture to arbitrate between competing goals of UI control algorithms in AR (Julier et al. 2003). This assertion was based on an analysis of operations that UI elements might wish to perform: suppress the display of an object, require the display of an object, or alter the on-screen representation of an object. Some operations happened in 2D and some in 3D. Based on potential conflicts, we argued for this more complex architecture than the pipeline we had previously used.

Our new filtering algorithm abandons these complex concepts, which had been proving difficult to implement and control. Instead, the new algorithm concentrates on fewer factors directly related to mobile military AR applications. These important factors include an operation zone that is defined in the planning stage and may be modified during the mission, the user, and objects that have impacts on the operation zone and the user. This enabled us to return to the much simpler pipeline architecture (Fig. 6). By incorporating the occlusion representation (Sect. 2.2) module into the information filter (Sect. 2.1), potential conflicts between these two features can be managed. The potential for conflict is reduced largely by reducing the use of the suppression by the filter. Because the user's focus now includes larger regions that by definition include destinations and other potentially more distant objects of interest, the loss of this operation does not pose serious problems for the users. Also, the SMEs were much more concerned with visualizations of personnel (friendly forces and known enemy locations) and control measures (routes, phase lines for synchronization, areas of responsibility, and restricted fire areas) rather than geometry of the environment.

---
[2] http://www.lua.org.
[3] http://www.openscenegraph.org/projects/osg.
[4] http://www.delta3d.org.
[5] http://qt.nokia.com/products.

*3.2.2 Interaction modes*

Our interactions now occur in two modes. One is through the map view, as described above—a purely 2D operating mode. The other is through direct 3D interaction. Our SMEs noted the use of ranging devices in military operations and suggested the inclusion of one in the system into which prototype UI is envisioned to be incorporated. This simplifies some interaction, as we can now specify 3D locations (at least those that are visible to the user) with 3D input (2D cursor plus depth). Further, we have automated the interaction module's needs to suppress or require the on-screen representation. This decision is now much more static, based on the user's focus. Although this may in theory change during an operation, our SMEs do not foresee frequent changes or the need to change while moving through the environment. Thus, we have restricted our UI pipeline to an ordered set of 3D operations followed by potential 2D operations. Though we leave as future work re-implementing the error adaptation and label placement algorithms described as part of the past architecture, we can see based on the revised analysis here (combined with the unaltered portions of the previous analysis) that this architecture will work with a pipeline.

*3.2.3 Visualizations*

As noted above, our SMEs surprised us by recommending that we not count on having a world model. This heavily affects several aspects of an AR system, not the least of which is the assumptions a video-assisted tracking system might make. But it also affected the representations we used in the occlusion representations described above. We restricted ourselves to methods that required at most a sparse world model, knowing that little could be done with more than a single real surface without a rough world model. We were able to identify several potential designs that appear promising under this restriction, however. Further user studies will aim to find a best representation for our mobile application. The C2 application represents a somewhat different problem, however, and we may find that a different method might be appropriate for a C2 user than what is best for the mobile user. Note that the C2 application does make the assumption of a sparse 3D world model, and this user requires a more global view of the actions occurring in the world.

We learned from interface design reviews with our SMEs that filled shapes were not favored for this SA application. It was felt that the coloring of interior regions would potentially interfere with the users' abilities to see the real environment, which is always a requirement. Thus, we restricted ourselves to line drawings for the representations of occluding surfaces. This in turn required modifications (described above) to some proposed occlusion metaphors.

The map view, a long-standing component of mobile AR applications, is very popular with military users; it is among the most familiar analogies our system can make with existing military equipment, either a paper or electronic map. Our SMEs expect such a view to assist in user acceptance of the prototype system. The use of military symbology also reflects this need for our system.

## 4 Conclusions

We have refined the UI architecture for our mobile application, finding a way to merge diverse aspects and features from previous AR implementations. Our development process relies on interviews with SMEs to help ensure that our implementations align with the needs and requirements of military personnel, but also that we are able to take advantage of results in UI research. Our integrated architecture was made possible by a more refined understanding of the tasks of users, which resulted in some simplifying assumptions that reduced the complexities in the components of the UI. Though we must still implement some of the latter stages of the pipeline we envision, we have reason to believe that this simpler architecture than the previous proposal, based on a "mediator" module, will serve our users' needs better and be much more successful at producing usable information presentation and interaction methods.

Other future work will include follow-up user studies to the pilot test described here. We see two needs: understanding the techniques with promise for the mobile user, and identifying appropriate techniques for the C2 user. Other user studies will focus on the interaction with the map view. A key component of this application's success will be the ability to communicate with personnel who are not within line-of-sight contact. The map is envisioned to be a key method by which such interaction may occur. All such studies must also eventually be conducted with military personnel, which requires that the application be seen as valuable enough for such personnel to invest their time in volunteering in such studies. Finally, after-action review capabilities are an important tool to military personnel, and recording capabilities must be implemented in both applications. Playback will be a feature of the C2 application.

Our UI architecture has enabled us to take a significant step forward in simplifying the user interface, both from the programmer's and the user's perspective. This architecture thus brings us closer to realizing our goal of placing such systems in the hands of end users.

**Acknowledgments** This work was supported in part by the NRL Base Program and in part by DARPA under the ULTRA-Vis Program for work as a performing member of the Lockheed Martin team, contract # FA8650-09-C-7908.